\documentclass{article}
\usepackage{emulateapj,pstricks,apjfonts}

\def\alwaysmath#1{\ifmmode{#1}\else{$#1$}\fi}

\righthead{Pancino, et al.}
\lefthead{}

\slugcomment{Draft version of \today}
\slugcomment{Submitted to the Astrophysical Journal, December 18, 1999}

\begin{document}

\title{New Evidence for the Complex Structure of the
Red Giant Branch in $\omega$ Centauri\altaffilmark{1}}

\author{Elena Pancino\altaffilmark{2},
 Francesco R. Ferraro\altaffilmark{2,3},
	Michele Bellazzini\altaffilmark{3},
	Giampaolo Piotto\altaffilmark{4},
	and
	Manuela Zoccali\altaffilmark{4}}
%\author{E. Pancino\altaffilmark{2},
% F. R. Ferraro\altaffilmark{2,3},
%	M. Bellazzini\altaffilmark{3},
%	G. Piotto\altaffilmark{4},
%	and
%	M. Zoccali\altaffilmark{4}}

\altaffiltext{1}{Based on observations collected at ESO, La Silla, Chile}
\altaffiltext{2}{European Southern Observatory, Karl Schwarzschild Strasse 2,
D-85748 Garching bei M\"unchen, Germany; epancino@eso.org, fferraro@eso.org}
\altaffiltext{3}{
Osservatorio Astronomico di Bologna, via Ranzani 1, I-40126
Bologna, Italy;
ferraro@bo.astro.it,
bellazzini@bo.astro.it.}
 \altaffiltext{4}{Dipartimento di Astronomia --- Universit\'a di Padova,
Vicolo dell'Osservatorio 5, I-35122 Padova, Italy; piotto@pd.astro.it,
zoccali@pd.astro.it}

\begin{abstract}

We report on the complex structure of the red giant branch (RGB) of
$\omega$ Cen, based on a new wide field and wide color baseline $B$
and $I$ photometry. Our color magnitude diagram (CMD) shows the
presence of multiple populations along this branch, in particular we
discovered an anomalous branch (RGB-a), which appears to be well
separated from the bulk of the RGB stars. On the basis of our CMD and
from the previous literature we conclude that (1) these stars, clearly
identified as a separate population in our CMD, represent the extreme
metal rich extension ($[Ca/H]>-0.3$) of the stellar content of
$\omega$ Cen, and show anomalous abundances of s-process elements (as
$Ba$ and $Zr$) as well; (2) they are physical members of the $\omega$
Cen system; (3) they comprise $\sim 5\%$ of the stars of the whole
system; (4) this component and the metal-intermediate one
($-0.4>[Ca/H]>-1$) have been found to share the same spatial
distribution, both of them differing significantly from the most metal
poor one ($[Ca/H]<-1$). This last evidence supports the hypothesis
that metal rich components could belong to an independent (proto?) stellar
system captured in the past by $\omega$ Cen.

\end{abstract}


\section{Introduction} \label{sec:intro}

The globular cluster
$\omega$ Centauri (NGC~5139) is the most luminous and massive object
among the Galactic Globular Clusters (GGC), and surely the most
peculiar one in terms of structure, kinematics and stellar content. It
is the most flattened GGC, displaying also a decrease of
ellipticity in the most internal region (Geyer, Nelles \& Hopp 1983 -
GNH83), and it has a significant rotation (Merrit, Meylan \& Mayor
1997).

The most interesting anomaly is its chemical inhomogeneity
(first revealed by Dickens and Woolley 1967 and spectroscopically
confirmed by Freeman \& Rodgers 1975); since then a number
of extensive spectroscopic surveys (Norris, Freeman \& Mighell 1996 -
NFM96, Suntzeff \& Kraft 1996 - SK96) have shown that $\omega$ Cen is
the only GGC for which a multi-component heavy element distribution
has been identified. Although SK96 found a single peaked distribution
with an extended tail towards high metallicities, NFM96 showed that the
distribution is at least bimodal with a main metal poor component at
$[Fe/H]\simeq -1.6$, a second smaller peak at $[Fe/H]\simeq -1.2$, and
a long tail extending up to $[Fe/H]\simeq -0.5$. Furthermore, the most
metal rich stars ($[Ca/H]>\sim -1.0$) have been found to be more
centrally concentrated than the bulk of the cluster population (Norris
et al. 1997 - N97, and SK96). It has also been suggested that the
kinematical properties (N97) and the spatial distribution of the two
metallicity groups differ significantly (N97, Jurcsik 1998 -
J98). This puzzling scenario has usually been explained either in
terms of self-enrichment processes (Freeman 1993) and/or merging
events (N97, J98).

As part of a long term project specifically devoted to the study of
the global stellar population in a sample of GGCs, we obtained wide
field $B$, $I$ photometry in $\omega$
Cen. The complete data set will be presented in a forthcoming paper,
while in this letter we concentrate on the complex structure of the
RGB.

\begin{figure*}[htb]
\vskip4truein
\includegraphics{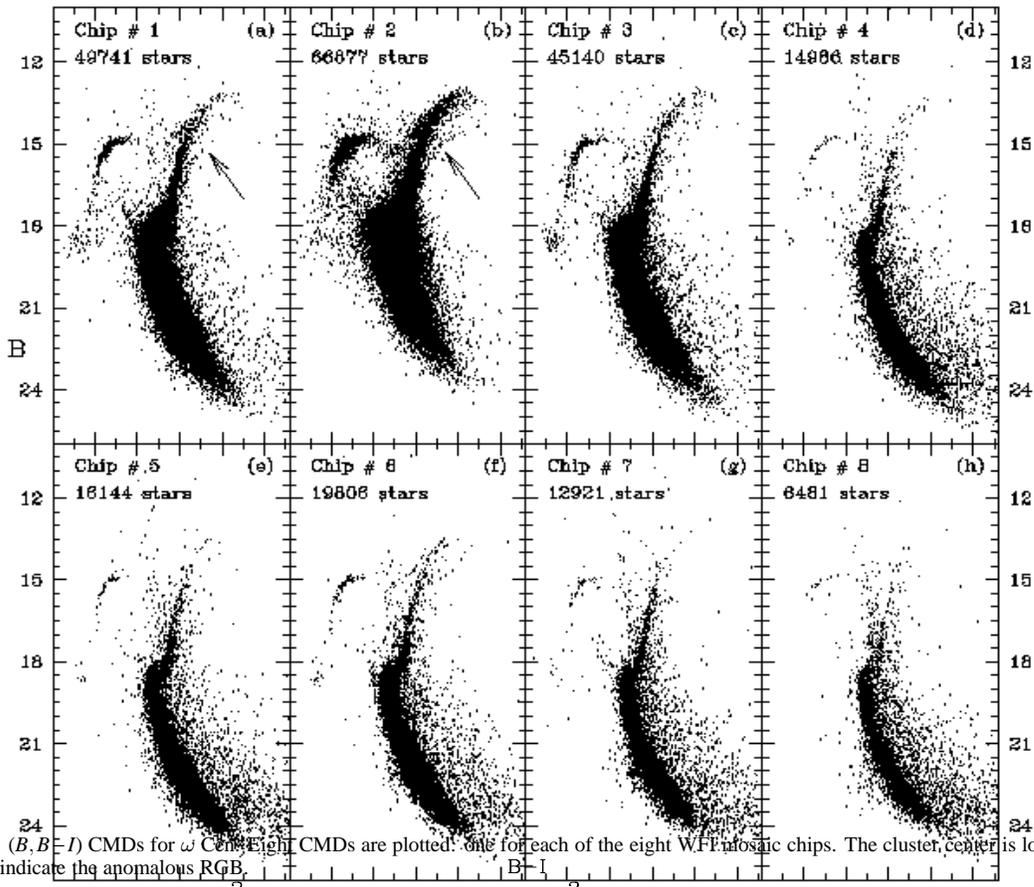}
\caption[fig1.ps]{
$(B,B-I)$ CMDs for $\omega$ Cen. Eight CMDs are plotted: one for each
of the eight WFI mosaic chips. The cluster center is located in Chip
\#2 ({\it panel (b)}). The arrows indicate the anomalous RGB.
\label{fig:map}}
\end{figure*}

\section{The RGB of $\omega$ Cen}
  
The data have been obtained on January 1999 and July 1999 at the 2.2m
ESO-MPI telescope at La Silla (Chile), using the Wide Field Imager
(WFI) which has a total field of view of $\sim 34'\times 33'$. The
images were taken through the $B$ filter and $I_{853}$, a Medium Band
Filter (cf. WFI@2.2 Manual) which avoids the most pronounced
atmospheric emission lines. After applying the standard bias and flat
field correction, we used DAOPHOT II and the psf-fitting algorithm
ALLSTAR (Stetson 1994) for the stellar photometry. The photometric
calibration was performed using 50 photoelectric standard stars in the
selected areas SA98 and SA95 (Landolt 1992).

\begin{figure*}[htb]
\vskip4truein
\includegraphics{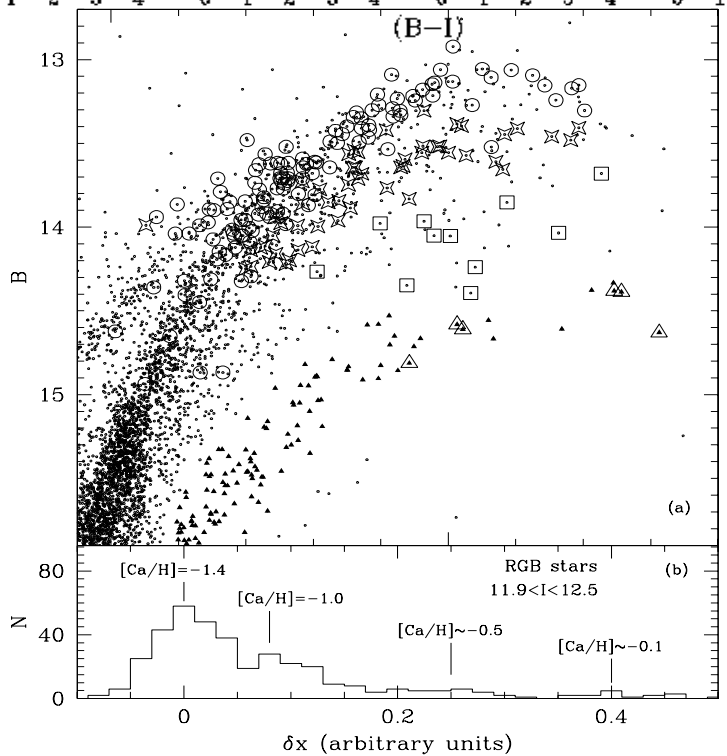}
\caption[fig2.ps]{
{\it Panel (a):} Zoomed CMD of the upper RGB region. Stars in the
RGB-a have been plotted as {\it small filled triangles}. Large empty
symbols show the position in the CMD for stars with known metallicity
(from NFM96). Different symbols refer to stars with different
metallicities: {\it large open circles} for stars with
$-1.5<[Ca/H]<-1.3$, {\it large open stars} for $-1.1<[Ca/H]<-0.85$,
{\it large open squares} for $-0.65<[Ca/H]<-0.4$, {\it large open
triangles} for $[Ca/H]>-0.3$, respectively.  {\it Panel (b):}
Histogram of the distribution of the distances from the MP-MRL (see
text). The mean $[Ca/H]$ abundances for the four main components of
the RGB ({\it panel (a)}) are also marked.
\label{fig:map}}
\end{figure*}

Figure 1 shows the $(B,B-I)$ CMDs for each of the 8 WFI chips
separately. The cluster center is roughly centered on chip $\#2$.
More than 220,000 stars have been plotted in Figure 1: to our
knowledge, this is the largest stellar sample ever observed in
$\omega$ Cen. The most striking feature of the CMDs presented in
Figure 1 is the existence of a {\it complex} structure in the brighter
part of the Red Giant Branch (RGB): at least two main components are
visible.

Particularly notable is the presence of a narrow sequence,
significantly redder and more bent than the bulk of the ``main'' RGB
stars, which we call the {\it anomalous} RGB (hereafter RGB-a). Note
that the RGB-a is visible only in the CMD from chip \#2 (where the
cluster center is located) and to a much lesser extent in the nearby
chip \#1 and possibly \#3 (see the arrows in Figure 1(a),(b)). We
shall come back to the spatial distribution of the RGB stars in the
next Section.  While we are writing this Letter, a $(V,B-V)$ CMD
showing the same RGB structure has been published by Lee et
al. (1999): the reality of this feature is thus confirmed by two
independent surveys based on  different photometric systems.

The morphology of the RGB-a and its position in the CMD strongly
suggests that it must be populated by stars much more metal rich than
the $\omega$ Cen bulk population. We can directly check this
hypothesis using the $[Ca/H]$ catalog by NFM96 (kindly provided by
J. Norris). We have cross-identified the stars in our catalog and in
the NFM96 one, and marked the common stars in Figure 2 ({\it panel (a)})
using different symbols for different metallicities (see figure
caption).

Six stars belonging to the RGB-a are in common with the NFM96 sample
(namely stars $ROA 300$, $ROA 447$, $ROA 500$, $ROA 512$, $ROA 517$,
$ROA 523$, adopting the Woolley 1966 nomenclature): these stars have
$[Ca/H]=-0.1\pm0.1$.  They are the most metal rich stars in the NFM96
catalog, though they do not correspond to any of the peaks shown in
the NFM96 $[Ca/H]$ distribution (e.g.  their Figure 12). Figure 2
suggests that the RGB-a represents an {\it additional}, very metal
rich component, located at the extreme tail of the abundance
distribution in $\omega$ Cen. Moreover, the large color baseline of
the present survey allows us to clearly separate the most metal rich
component from the remaining $\omega$ Cen RGB stars and, thus, to
estimate its contribution to the {\it global} population of the
cluster. By comparing the star counts along the RGBs, in the same
magnitude range ($B<16$), we find that the RGB-a represents $\sim 5\%$
of the whole stellar content. It is important to note that the six
RGB-a stars for which we know the radial velocities from the Mayor et
al. (1997) catalog (kindly provided by G. Meylan) have $v_{rad}$ =
$233\pm9~km~s^{-1}$. This value is fully compatible with the mean
value $<v_{rad}>=233\pm17~km~s^{-1}$ for the 471 stars in the Mayor et
al. (1997) catalog, showing that the RGB-a stars are indeed members of
the $\omega$ Cen system (see also Lloyd Evans, 1983 - LE83).
 
To further investigate the distribution of the RGB stars in the CMD,
we computed the distance of each of them from the mean ridge line of
the main metal poor component (MP-MRL) in several magnitude bins.  The
MP-MRL has been obtained following the procedure described in Ferraro
et al. (1999a). A preliminary selection of the stars belonging to the
bluest ridge of the RGB (excluding HB, AGB, and the reddest RGB
components) has initially been performed by eye. Then we fitted the
selected stars with polynomials, and the procedure was iterated (each
time, rejecting all stars at $>2 \sigma$ from the best fit line),
until the result was considered acceptable. In doing this, we used the
$(I,B-I)$ CMD, since in this plane the upper part of the RGB is less
bent than in $(B,B-I)$, and a low order polynomial excellently
reproduces the RGB shape.

Then we defined the observable $\delta x$ as the geometrical distance
of each star from the adopted MP-MRL. The distribution of $\delta x$
for the RGB stars in the magnitude range $11.9<I<12.5$ is plotted in
Figure 2 ({\it panel (b)}). Positive and negative values of $\delta x$
are assumed for stars redder and bluer than the MP-MRL,
respectively. The distribution is clearly asymmetric; besides the main
peak, secondary peaks are well visible. The estimated typical $[Ca/H]$
abundances for the different components [as in {\it panel (a)}] are
marked in {\it panel (b)}. Four different components can be identified
in Figure 2. They can be grouped in three main samples as
follows: ({\it i}) {\bf RGB-MP} for the most metal poor component, at
$[Ca/H]\sim -1.4$; ({\it ii}) {\bf RGB-MInt} for the metal
intermediate component: this sample includes the secondary peak
($\delta x \sim 0.08$) in Figure 2(b), which corresponds to
$[Ca/H]\sim -1.0$, and the metal rich tail ($\delta x \sim 0.2-0.3$),
corresponding to $[Ca/H]\sim -0.5$; ({\it iii}) {\bf RGB-a} for
the extreme metal rich component, at $\delta x \sim 0.4$ in Figure
2(b), i.e. $[Ca/H]\sim -0.05$, clearly separated from the
other stellar populations.  The first two components correspond to the
peaks visible in Figure 12 in NFM96, while the RGB-a is a third, new
component of the stellar population of $\omega$ Cen (see also Lee et
al. 1999).

To better describe the characteristics of the ``multiple'' $\omega$
Cen RGB in terms of the ``classical'' abundance parameter $[Fe/H]$, we
used a set of 60 giants in our sample which have both $[Ca/H]$ (from
NFM96) and $[Fe/H]$ abundances (from SK96, kindly provided by
N. Suntzeff). From these, we derived an empirical relation linking the
calcium abundance to iron, in terms of $[Fe/H]$ on the Zinn \& West
(1984) scale. The procedure yields the following results for the four
peaks labeled in Figure 2 ({\it panel(b)}): $[Fe/H]_{ZW}\sim -1.7,
-1.3, -0.8$ and $-0.4$, respectively.
  
It is interesting to note that the spectra of all of the six metal
rich stars quoted above exhibit strong $BaII$ lines, and the brightest
three (namely $ROA300$, $ROA447$ and $ROA513$) show also strong $ZrO$
bands (LE83), a very unusual occurrence for globular cluster giants.
The fact that high metallicity giants have strong $Ba$ lines fully
agrees with the increase of the $[Ba/Fe]$ ratio with metallicity found
by Norris \& Da Costa (1995), and suggests that intermediate mass ($M
< 10 ~M_{\odot}$) stars may have contributed significantly to the
enrichment of the material from which RGB-a stars formed
\footnote{Both $Ba$ and $Zr$ are elements produced by neutron 
{\it s}-capture processes. {\it s}-process elements are brought up to
the surface during the third dredge up occurring at late stages of the
asymptotic giant branch (AGB) evolutionary phase of intermediate mass
stars (see the discussion in Smith, Cunha \& Lambert 1995). However
LE83 noted that {\it s}-element rich giants in $\omega$ Cen show
different characteristics with respect to normal AGB stars and first
suggested that the anomalous {\it s}-elements enrichment could be
primordial.}.  This, in turn, implies a significant delay ($\sim 0.5
Gyr$) between the formation of the stellar population that enriched
the medium and the one associated with RGB-a (Norris \& Da Costa 1995,
Lee et al. 1999).
  
\section{Spatial distribution anomalies}
 
The correlation between metal content and kinematics found by N97, and
the spatial asymmetry between metal rich and metal poor bright red
giants found by J98, are a possible evidence of a past merging event,
in which a small metal rich object has been captured by a larger
system, mainly metal poor. Note that none of the quoted studies have
identified the RGB-a as a distinct population, since only a few stars
belonging to this branch have been measured by NFM96 and listed by
J98. The discovery of a population so clearly separated from the main
branches makes the hypothesis that $\omega$ Cen hosts an external
population captured in the past more and more appealing, even if it is
worth noticing, according to note 3 by N97, that neither the large
metal spread nor the s-process abundances can be easily explained in
terms of a simple merging event between ordinary globular clusters.
 
These considerations prompted us to investigate the spatial
distribution of the three RGB components, taking advantage of our
photometric sample, which is much larger ($N_{RGB}>3500$) and more
complete than the N97 and J98 ones.

As a first step, we evaluated the centroid of the stars of the entire
sample; by adopting the procedure described in Montegriffo et
al. (1995) and Ferraro et al. (1999b), we found it to be at pixel
$(700\pm20,1900\pm20)$. The centroids of all three of the RGB
components appear to coincide within the observational errors.

Then, we investigated the radial distributions; special care has been
devoted to avoid contamination by the field population which can
affect the radial trend, expecially in the lower density external
region. To this purpose, we limited our analysis to the most internal
region ($r<10'$), and to the brighter portion ($B<15.5$) of the RGB,
which is expected to be little contaminated. The comparison of the
cumulative radial distributions for the selected RGB samples shows
that the RGB-MInt and RGB-a stars have a very similar distribution,
and both seem to be more concentrated than the RGB-MP. A KS test gives
an $8\%$ probability that the radial distribution of the coadded
(RGB-MInt $+$ RGB-a) sample and the RGB-MP sample are drawn from the
same parent distribution. Although this evidence is marginal, it is in
good agreement with that found by N97. Still, this simple radial
distribution analysis implicitly assumes that the three populations
are distributed with the same (spherical) geometry. In view of the
strong ellipticity of $\omega$ Cen, we decided to follow an
alternative approach.

In order to avoid the sparsely populated outer regions, we divided the
central region of the cluster ($r<13'$ from the center) into a grid of
$13\times 13$ square boxes, each $500\times 500$ pixels wide. Inside
each $(i,j)-th$ box, we counted the number of stars belonging to the
different samples ($N_{RGB-MP}^{ij}$, $N_{RGB-MInt}^{ij}$ and
$N_{RGB-a}^{ij}$, respectively), thus obtaining a density map for each
component. The resulting isodensity contour maps, down to a fixed
density limit (4 stars per box), are shown in Fig. 3. The
distributions of the RGB-a sample (heavy solid lines in {\it panel
(a)}) and of the RGB-MInt (heavy solid lines in {\it panel (b)}) are
compared with the RGB-MP distribution (light dashed lines in both
panels). The elongation of the RGB-MP isopleth along the X-axis is
clearly visible, reflecting the well known elliptical shape of the
whole system: our X-axis is approximately oriented along the E-W
direction, i.e., the major axis of the cluster. It is worth noticing
that the ellipticity for the RGB-MP component turns out to be
$\epsilon \sim 0.2$, even in the most internal region of the cluster
($2'-4'$). On the other hand, both the RGB-a and RGB-MInt isopleth
show a very different direction of maximum elongation, nearly
perpendicular to the RGB-MP one (i.e., along the N-S
direction). Furthermore, a very peculiar substructure is visible in
the RGB-a contours (see {\it panel(a)}), resembling the tidal tails
which should be expected to form around a disrupting stellar system
(e.g., Meylan, Leon \& Combes 1999).

\begin{figure*}[htb]
\vskip7.5truein
\includegraphics{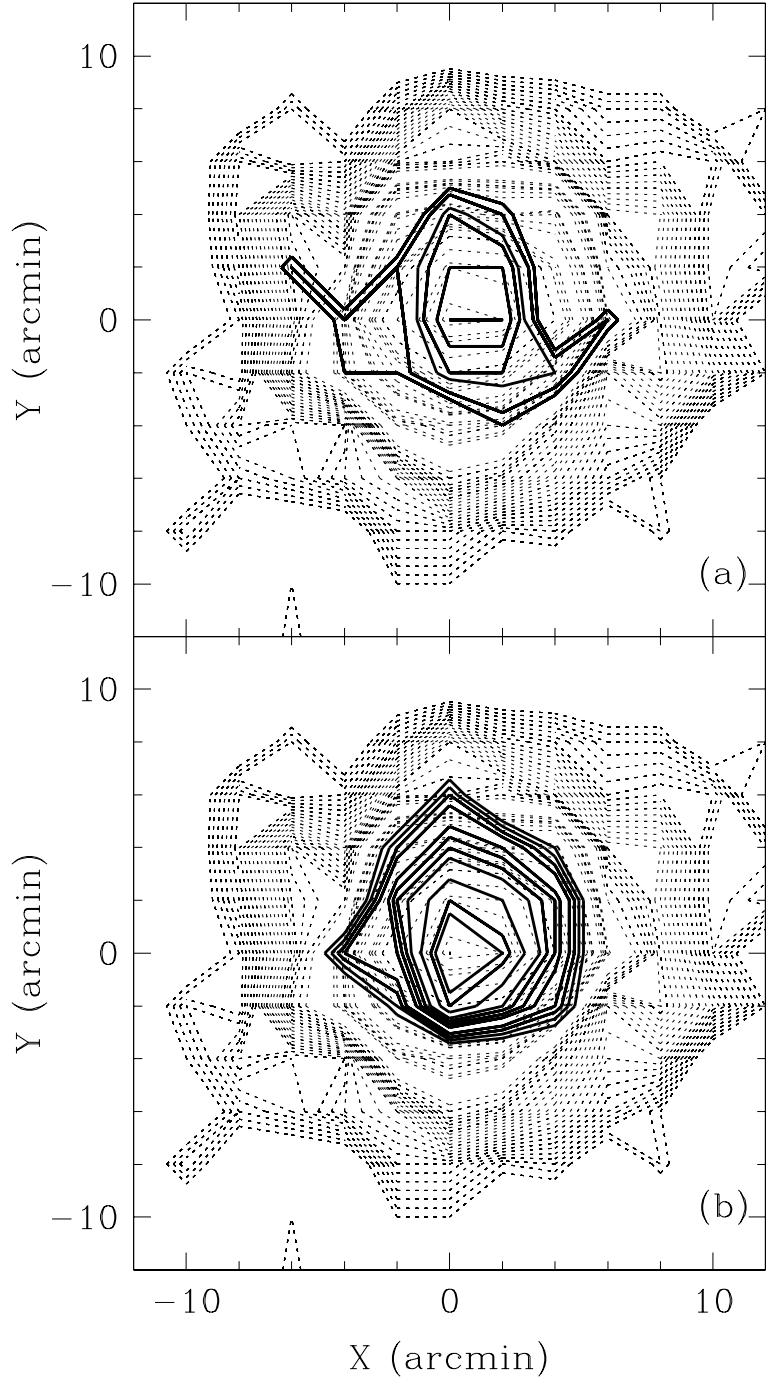}
\caption[fig3.ps]{
The isodensity contour maps. The map is $\sim 26'$ in size.
The distribution of the RGB-a sample (heavy solid line in {\it panel
(a)}) and of the RGB-MInt (heavy solid line in {\it panel (b)}) is
compared with the RGB-MP distribution (light dashed lines in both
panels). The step between the isopleths is 2 counts for the RGB-a and
RGB-MInt maps. For the RGB-MP the step is 1 count for the outer 30
isopleths and 10 for the others. The total number of stars comprised
in the outer isopleth is $N=116$ for the RGB-a, $N=236$ for the
RGB-MInt and $N>1000$ for the RGB-MP samples. The difference in the
elongation of the spatial distribution between the RGB-MP (along the
X-axis) and the RGB-MInt, RGB-a (along the Y-axis) is well visible. 
\label{fig:map}}
\end{figure*}

We took advantage of the fact that the major axis of the RGB-MP
ellipsoid is nearly oriented along our X-axis, while the RGB-MInt and
RGB-a ones are nearly oriented along the Y-axis. We can thus study the
distribution along the two main symmetry axes by simply comparing the
distributions along our X and Y coordinates with respect to the the
common centroid. The main results of this analysis are: (1) the RGB-MP
sample is significantly more concentrated in the Y direction than in X
(confidence level $99.5\%$, according to a KS test), as expected for
an ellipsoid elongated in the X-axis direction; (2) the RGB-MInt and
RGB-a do not share the elongation in the X direction of the RGB-MP
stars, and look more elongated in Y than the RGB-MP ones
(c.l. $99.6\%$); (3) Fig. 3 ({\it panel (b)}) suggests that the
RGB-MInt stars distribution is strongly asymmetric along the Y-axis,
being more elongated towards the North direction.

These results could help to shed some light on previous puzzling
evidence. In particular: point (1) and (2) could explain the decrease
of ellipticity in the inner $5^{\prime}$ of the cluster reported by
GNH83, and point (3) could help to understand the bizarre spatial
asymmetry found by J98.

Furthermore, the similarity in the spatial distributions may indicate
a physical association between the RGB-a and RGB-MInt, while the
differences shown in Figure 3 suggest a different origin of these two
components with respect to the main metal poor one (RGB-MP). If this
result finds additional support, and the primordial origin of the
s-elements found in the RGB-a stars is definitely established, then
the RGB-a and RGB-MInt populations can be interpreted as two
successive phases of the self-enrichment history of an independent
stellar system (a giant molecular cloud or a gas rich protocluster),
now sunk into the center of $\omega$ Cen.

In conclusion, we can say that the observational evidence discussed in
this Letter seem to suggest that both self-enrichment processes and a
merging event should be invoked to explain the complex structure of
the RGB in $\omega$ Cen.

\acknowledgments

We thank N. Suntzeff, R. Kraft, J. Norris, G. Meylan and J. Jurcsik
for providing their data in computer readable form, and Luca Pasquini
and Alvio Renzini for many stimulating discussions. We are also
indebted to an anonymous referee for the helpful suggestions on the Ba
and ZrO anomalies and to Tad Pryor that referred the paper for what
concerns dynamics for his detailed report. 
The financial support of the {\it
Ministero della Universit\`a e della Ricerca Scientifica e
Tecnologica} (MURST) to the project {\it Stellar Dynamics and Stellar
Evolution in Globular Clusters} and to the project {\it Treatment of
large-format astronomical images} is kindly acknowledged. EP thanks
the {\it ESO Imaging Survey Visitor Program} and the EIS team for
providing technical support. FRF gratefully acknowledges the
hospitality of the {\it Visitor Program} during his stay at ESO, when
most of this work has been carried out.

\end{document}